\documentclass{article}
\usepackage[utf8]{inputenc}
\usepackage[a4paper, total={6in, 8in}]{geometry}
\usepackage{graphicx}
\usepackage{xcolor}
\usepackage{setspace}
\usepackage{multicol}
\usepackage{mathtools, nccmath}
\usepackage{booktabs}
\usepackage{tabularx}
\usepackage{authblk}
\usepackage{hyperref}  %
\usepackage{natbib}
\setcitestyle{numbers,square}

\title{Quantum metric non-linear Hall effect in an antiferromagnetic topological insulator thin-film EuSn$_2$As$_2$}

\author[1]{Hung-Ju Tien}
\author[2]{Hsin Lin}
\author[3]{Liang Fu}
\author[1,4,5]{Tay-Rong Chang\thanks{u32trc00@phys.ncku.edu.tw}}
\affil[1]{Department of Physics, National Cheng Kung University, Tainan 70101, Taiwan}
\affil[2]{Institute of Physics, Academia Sinica, Taipei 115201, Taiwan}
\affil[3]{Department of Physics, Massachusetts Institute of Technology, Cambridge, MA 02139, USA}
\affil[4]{Center for Quantum Frontiers of Research and Technology (QFort), Tainan 70101, Taiwan}
\affil[5]{Physics Division, National Center for Theoretical Sciences, Taipei 10617, Taiwan}

\begin{document}

\maketitle

\begin{abstract}
The quantum geometric structure of electrons introduces fundamental insights into understanding quantum effects in materials. One notable manifestation is the non-linear Hall effect (NLHE), which has drawn considerable interest for its potential to overcome the intrinsic limitations of semiconductor diodes at low input power and high frequency. In this study, we investigate NLHE stemming from the real part of the quantum geometric tensor, specifically the quantum metric, in an antiferromagnetic topological material, EuSn$_2$As$_2$, using density functional theory. Our calculations predict a remarkable NLHE arising from a symmetry-protected, single Type-II surface Dirac cone in the even-numbered-layer two-dimensional slab thin-film, yielding a non-linear Hall conductivity exceeding 20 mA/V$^2$—an order of magnitude larger than previously reported. This single Dirac band dispersion represents the simplest model for generating NLHE, positioning the EuSn$_2$As$_2$ thin-film as a “hydrogen atom” for NLHE systems. Additionally, we observe NLHE from band-edge states near the Fermi level. Our findings also reveal that 30$\%$ phosphorus (P) doping can double the non-linear Hall conductivity. With its substantial and tunable NLHE, EuSn$_2$As$_2$ thin-films present promising applications in antiferromagnetic spintronics and rectification devices.
\end{abstract}



\section{Introduction}

 First discovered by Edwin Hall in the 19$^{th}$ century, the Hall effect elucidates the generation of transverse voltage when both a longitudinal current and an external magnetic field are concurrently applied ~\cite{RN196}. Later, Hall identified that this transverse Hall voltage can manifest in ferromagnetic materials even without an external magnetic field, a phenomenon he termed the anomalous Hall effect ~\cite{RN197}. With advancements in research on the geometric phase of electronic systems, we now understand that the linear component of the anomalous Hall effect arises from the Berry curvature induced by the broken time-reversal symmetry in solid-state systems ~\cite{RN180},~\cite{RN122}.

The non-linear Hall effect (NLHE) has recently garnered global attention beyond the linear response of the anomalous Hall effect, sparked by experimental observations of a transverse non-linear Hall voltage even under conditions of time-reversal symmetry ~\cite{RN92},~\cite{RN93},~\cite{TRC_4} affirming the theoretical prediction linking NLHE to the Berry curvature dipole (BCD)  ~\cite{RN86}. Subsequent, numerous potential applications leveraging NLHE have been proposed, including terahertz detection and rectification ~\cite{RN177},~\cite{RN178},~\cite{RN165}, along with its utility as an indicator of topological transitions ~\cite{RN89}.

Recent research efforts have predominantly delved into exploring the NLHE within antiferromagnetic (AFM) systems that uphold the space-time reversal $(PT)$ symmetry ~\cite{RN87},~\cite{RN88},~\cite{RN120},~\cite{RN194},~\cite{RN164}. In such systems, the $PT$ symmetry mandates the vanishing of both the Berry curvature and the BCD across the entire Brillouin zone (BZ). Given the inability of BCD to induce NLHE under $PT$ symmetry, an alternative origin, the quantum metric dipole, has been proposed to evoke NLHE in AFM materials ~\cite{RN85}. Within this framework, the non-linear Hall conductivity can be expressed as:
\begin{equation}
        \begin{aligned}[b]
        \sigma^{\alpha\beta\gamma} = 2e^3\sum_{n,m}^{E_n \neq E_m}\int \frac{d^3k}{(2\pi)^3} 
        & \frac{v^\alpha_n g^{\beta\gamma}_n}{E_n-E_m}
        \frac{\partial{f(E_n)}}{\partial{E_n}} -
        (\alpha \leftrightarrow \beta)
        \end{aligned}
\end{equation}, where $n$ and $m$ denote indices for different bands, $E_n$ represents the energy of the Bloch state, $v^{\alpha}_n$ signifies the band velocity, and $g^{\beta\gamma}_n$ stands for the quantum metric tensor, the real part of quantum geometry ~\cite{RN174}. The quantum metric tensor $g^{\beta\gamma}_n$ measures the distance between neighboring Bloch states in Hilbert space, defined as $g^{\beta\gamma}_{n} = Re\sum_{i,j}^{n \neq m}[A_{ni,mj}^{\beta}A_{mj,ni}^\gamma]$, where ${\bf\it A}_{n,m} = i\langle u_n | \nabla_k | u_m \rangle$ denotes the Berry potential with the periodic part of the Bloch state ~\cite{RN87},~\cite{RN85}. Thus, $v^{\alpha}_{n}g^{\beta\gamma}_{n}$ can be interpreted as the quantum metric dipole. Remarkably, this Hall conductivity remains independent of relaxation time, originating from the inherent nature of electronic structure, termed the intrinsic non-linear Hall effect (INHE) ~\cite{RN87},~\cite{RN85}. Recent studies suggest utilizing INHE in rectification devices or generating DC currents within the WiFi frequency range ~\cite{RN165},~\cite{RN179}. Given that device efficiency hinges on the NLHE, the pursuit of materials exhibiting substantial NLHE becomes pivotal, not only in condensed matter physics but also in material sciences. According to Eq. (1), a large INHE can naturally be expected in band structures characterized by small gaps and large Fermi velocities. Moreover, tilted band dispersion is a crucial requisite to prevent quantum metric cancellation in the BZ ~\cite{RN87},~\cite{RN88}. Following this rationale, tilted Dirac surface states in topological materials with in-plane antiferromagnetic spin configuration emerge as promising candidates.

In this work, we propose utilizing the AFM topological insulator EuSn$_2$As$_2$ ~\cite{RN96} with an even-numbered layer slab thin-film structure as a promising candidate to realize the INHE induced by the quantum metric dipole. This thin-film configuration preserves the $PT$ symmetry, with an additional $k_x = 0$ mirror plane $M_{[100]}$ protecting a nearly gapless topological surface state on the (001) plane. Our calculations reveal a significant non-linear Hall conductivity around this Type-II surface Dirac point. Specifically, the non-linear Hall conductivity of EuSn$_2$As$_2$ exceeds 20 mA/V$^2$, surpassing previous findings by an order of magnitude (see Table 1). Furthermore, we demonstrate that the INHE of EuSn$_2$As$_2$ can be dynamically tuned by rotating the direction of the spin moment with a period of $2\pi$. This tunable characteristic holds promise for enhancing the read-out information of Neel vectors in AFM spintronics ~\cite{RN182}, ~\cite{RN183}, ~\cite{RN184}, ~\cite{RN185}. The single Dirac band dispersion serves as the simplest model to generate INHE. Hence, the even-numbered layer EuSn$_2$As$_2$ slab thin-film can be likened to the hydrogen atom of INHE. Additionally, apart from the INHE stemming from surface Dirac band dispersion, our calculations unveil other substantial non-linear Hall conductivity induced by anti-crossing states at the band-edge-states near the Fermi level. Moreover, introducing a 30 $\%$ doping of phosphorus (P) pushes the non-linear Hall conductivity to approximately $45$ mA/V$^2$, markedly enhancing the functionality of EuSn$_2$As$_2$.


\begin{table}[h]
    \centering
    \caption{Reported intrinsic nonlinear Hall conductivity values are presented using two unit conventions. The unit mA/V$^2$ is applied to 3D periodic systems and real 2D systems with finite thickness. In contrast, the unit nm$\cdot$mA/V$^2$ is used when treating the material as an ideal 2D system without considering its thickness.}
    \label{tab:table1}
    \begin{tabular}{ccc}
        \toprule
        \textbf{Material} & \textbf{$\sigma^{\alpha\beta\gamma}$}  & \textbf{System} \\
        \midrule
        CuMnAs~\cite{RN87}    &   1.5$^{cal}$ (mA/V$^2$)    & AFM-metal (3D) \\ 
        Mn$_2$Au ~\cite{RN88}  &  -0.012$^{cal}$ (mA/V$^2$)  & AFM-metal (3D) \\
        4SL-MnBi$_2$Te$_4$/BP~\cite{RN179} & 8$^{exp}$ (mA/V$^2$)  & AFM-TI (2D) \\
        Monolayer MnS~\cite{RN194}  &  5$^{cal}$ (nm$\cdot$mA/V$^2$)    & AFM (2D) \\
        Monolayer TaCoTe$_2$,~\cite{RN120} & 4$^{cal}$ (mA/V$^2$)  & AFM-insulator (2D) \\
        EuSn$_2$As$_2$ (this work)  & 22$^{cal}$ (mA/V$^2$) &  AFM-Dirac (2D) \\
                                    & 88$^{cal}$ (nm$\cdot$mA/V$^2$) & \\
        
        \bottomrule
    \end{tabular}
\end{table}

\section{Symmetry constraint of intrinsic non-linear Hall effect}

For the second-order response, the relationship between current and electric field is described by $J \sim \sigma E^2$, where $\sigma$, $J$, and $E$ represent the second-order non-linear conductivity, electrical current, and electric field, respectively. Since $E^2$ is $P$ and $T$ even and $J$ is $P$ and $T$ odd, the second-order Hall conductivity must be $P$ and/or $T$ odd, necessitating the breaking of spatial inversion and/or time-reversal symmetry within the system. In Eq. (1), $\sigma^{\alpha\beta\gamma}$ denotes the non-linear Hall conductivity, exhibiting antisymmetry under $\alpha \leftrightarrow \beta$ and symmetry under $\beta \leftrightarrow \gamma$ permutations. Thus, in two dimensions (2$D$), only two independent components of the non-linear Hall conductivity, $\sigma^{yxx}$ and $\sigma^{xyy}$, exist. To analyze the symmetry constraints imposed by point group operations, the conductivity tensor in a 2$D$ system can be transformed into an equivalent pseudovector form ~\cite{RN134}:
\begin{equation}
\sigma^{\gamma} \equiv \epsilon^{\alpha\beta}\sigma^{\alpha\beta\gamma}/2
\end{equation}.
Here, $\alpha$, $\beta$, and $\gamma$ are spatial indices, and $\epsilon^{\alpha\beta}$ denotes the 2$D$ Levi-Civita symbol. According to this transformation rule, $\sigma^{yxx}$ and $\sigma^{xyy}$ can be rewritten as $\sigma^{x}$ and $\sigma^{y}$ (pseudovectors along the $\hat{x}$ and $\hat{y}$ directions), respectively. The transformation of the pseudovector under point group symmetry operation is expressed as:
\begin{equation}
        \sigma^{n}  = \eta_{T}det(R)R^{n\gamma}\sigma^{\gamma}
\end{equation},
where $n$ represents the spatial index, $R$ signifies the point group operation, and $\eta_{T} = -1$ denotes the magnetic symmetry operations of the form $PT$ ~\cite{RN88}. Thus, the non-linear Hall conductivity induced by the quantum metric dipole is time-reversal odd. The direction of the nonzero pseudovector $\sigma^n$ in the 2$D$ system must be orthogonal to the crystalline mirror planes. Consequently, the non-linear conductivity is completely suppressed in a 2$D$ material with two or more mirror planes, thereby establishing one mirror plane as the highest crystalline reflection symmetry permitted in a 2$D$ system.

\section{Crystal structures}

The crystal structure of EuSn$_2$As$_2$ adopts a rhombohedral lattice akin to that of Bi$_2$Se$_3$, where Eu atoms interconnect with the Sn-As hexagonal network to form hexagonal layers that stack along the $c$ direction. The space group of EuSn$_2$As$_2$ is No. 166. Previous studies have identified the magnetic ground state of EuSn$_2$As$_2$ as an in-plane A-type antiferromagnet ~\cite{RN96}. The band structure calculations reveal that EuSn$_2$As$_2$ preserves multiple topological invariant numbers ~\cite{RN96}, ~\cite{TRC_5}. The side surface exhibits $S$-symmetry-protected topological surface states. Additionally, a mirror-symmetry-protected gapless Dirac surface state is present away from $\bar{\Gamma}$ on the (001) surface, as confirmed by ARPES ~\cite{RN96}, ~\cite{TRC_5}. Due to the bulk structure's centrosymmetry, second-order conductivity is forbidden by symmetry constraints. By contrast, a finite-size 2$D$ slab structure breaks this centrosymmetry. We define the number of layers in the EuSn$_2$As$2$ thin-film by counting the Eu layers in the slab model. For instance, Fig. 1(a) and Fig. 1(b) depict 5-layer and 4-layer slab thin-films in our definition, respectively. Notably, an odd-number of layers maintains inversion symmetry akin to the bulk structure, while an even-number of layers breaks inversion symmetry. The remaining symmetry operators in the even-numbered layer slab geometry of EuSn$_2$As$_2$ are $PT$, $C_{2[100]}T$, and $M_{[100]}$. Adhering to this symmetry constraint, only $\sigma^{yxx}$ (pseudovector along the $\hat{x}$ direction) behaves as an even function under $M_{[100]}$ (Fig. 1(d)). Consequently, as depicted in the schematic in Fig. 1(e), a non-linear Hall voltage is anticipated to emerge along the $\hat{y}$ axis in response to an external electric field along $\hat{x}$.

\section{Computational method}

    The bulk band structure calculations for EuSn$_2$As$_2$ were conducted employing the projected augmented wave method as implemented in the VASP package, utilizing the generalized gradient approximation (GGA) ~\cite{RN187},~\cite{RN188} and GGA plus Hubbard $U$ (GGA+$U$) scheme ~\cite{RN190}. For the Eu 4-$f$ orbitals, an on-site $U$ value of 5 eV was employed. Experimental structural parameters were adopted ~\cite{RN96},~\cite{RN72}. Spin-orbit coupling (SOC) was self-consistently included in the calculations using a Monkhorst-Pack k-point mesh of 23 $\times$ 23 $\times$ 3. Wannier functions were constructed from Eu $d$, $f$, Sn $p$, and As $p$ orbitals without the need for the maximization of localization procedure ~\cite{RN100},~\cite{RN189}.

\section{Intrinsic non-linear Hall effects}

Figure 2(a) depicts the band structure of a 4-layer slab model of EuSn$_2$As$_2$ exhibiting in-plane A-type antiferromagnetism. We symmetrically construct a finite-size slab model terminated with As atoms. The band dispersion reveals a mirror symmetry protected fourfold-degenerate Dirac point on (001) along $\overline{M}-\overline{\Gamma}$, denoted as DP in Fig. 2(a). An enlarged view of the band structure (inset of Fig. 2(a)) reveals a Type-II Dirac band dispersion formed by the crossing between the conduction and valence bands. This band feature resembles the topological surface state of bulk EuSn$_2$As$_2$ ~\cite{RN96}, suggesting that the 4-layer slab thickness is sufficient to exhibit surface Dirac band dispersion (see Supplementary material). The corresponding intrinsic non-linear Hall conductivity $\sigma^{yxx}$ is illustrated in Fig. 2(b). Our calculations unveil three significant peaks around the Fermi level, labeled $\alpha$, $\beta$, and $\gamma$. Notably, the maximum conductivity exceeds 20 mA/V$^2$ at 20 K, which is an order of magnitude larger than previous studies (see Table 1), suggesting EuSn$_2$As$_2$ thin-film as a promising candidate for enhancing the operating efficiency of rectification devices. Non-linear Hall conductivity at different temperatures is presented in the Supplementary material.

To elucidate the origin of the INHE in EuSn$_2$As$_2$, we plot the intensity of the quantum metric dipole $D_{qm}=v_yg_{xx}-v_xg_{yx}$ on the band structure, as depicted in Fig. 2(a). Our calculations reveal high intensity of $D_{qm}$ around the surface Dirac point. Owing to the characteristics of the Type-II Dirac cone, the large slope of band dispersion and small energy gap simultaneously augment and diminish the Fermi velocity $v^{\alpha}_n$ and energy gap $(E_n-E_m)$ in Eq. (1), respectively. These concurrent changes amplify the quantum metric $g^{\beta\gamma}_n$, resulting in a pronounced intensity quantum metric dipole $D_{qm}$ around the Dirac point ~\cite{RN88}. Further, to identify the source of peak $\alpha$, we plot the constant energy contour at the energy of peak $\alpha$ and the intensity of $D_{qm}$ in Fig. 2(d). Our calculations indicate that the Dirac point is the sole hot spot, with no comparable contribution elsewhere in the BZ, establishing that this significant value of INHE stems entirely from the topological surface state. Crucially, the 2$D$ Dirac band dispersion represents the simplest model for generating INHE. Therefore, EuSn$_2$As$_2$ thin-film can be likened to the hydrogen atoms of INHE, providing a practical material platform for studying INHE and related properties. Additionally, the effects of strain and thickness on the INHE are discussed in detail in the Supplementary Material.

Given the high sensitivity of the surface Dirac cone and non-linear Hall conductivity of EuSn$_2$As$_2$ to crystalline symmetry, altering the direction of the spin moment presents a viable strategy for controlling the INHE. Figure 3 illustrates the non-linear Hall conductivity $\sigma^{yxx}$ and $\sigma^{xyy}$ originating from the surface Dirac cone with different in-plane spin orientations. Here, $\theta$ in Fig. 3 denotes the angle between the spin moment vector and the $x$ axis. Notably, the non-linear Hall conductivity follows the relationship $\sigma(\theta) = -\sigma(\theta+\pi)$, demonstrating a periodicity of 2$\pi$. Specifically, we observe that $\sigma^{yxx}$ achieves its maximum value while $\sigma^{xyy}$ remains at 0 when $\theta$ = 0 (spin moment aligned with the $x$ axis), owing to the enforcement of $M_{[100]}$ causing $\sigma^{xyy}$ to vanish as per Eq. (3). The illustration of the mirror plane with spin configuration is depicted in Fig. 1(c). As the spin direction deviates from the $x$ axis, the protection of the surface Dirac cone by $M_{[100]}$ is compromised, leading to gap opening and a decrease in $\sigma^{yxx}$ intensity. Conversely, with the destruction of $M_{[100]}$, $\sigma^{xyy}$ is no longer restricted, resulting in an increase in its value. At $\theta = \pi/2$, $\sigma^{yxx}$ vanishes due to enforcement by $M_{[120]}$, while $\sigma^{xyy}$ exhibits significant non-linear Hall conductivity. This tunability not only facilitates the detection of EuSn$_2$As$_2$ symmetry but also holds potential for advancing spintronics, such as measuring the Neel vector in AFM systems ~\cite{RN87}~\cite{RN88}.

In addition to the $\alpha$ peak in Fig. 2(b) originating from the surface Dirac cone, we observe two additional peaks, $\beta$ and $\gamma$, whose values are comparable to $\alpha$, despite their energies being distant from the Dirac point. To delve into the origins of these two peaks, we plot the constant energy contour with the intensity of $D_{qm}$ corresponding to $\beta$ and $\gamma$ as depicted in Fig. 2(e) and 2(f), respectively. From Fig. 2(a) and 2(e), we discern that the $\beta$ peak is contributed by the $D_{qm}$ accumulated along the $\overline{\Gamma}-\overline{K}$ direction. This heightened $D_{qm}$ strength arises from the small band gap induced by the anti-crossing of band-edge-states, denoted as BES Fig. 2(a). Similarly, the origin of the $\gamma$ peak also stems from the intricate anti-crossing of band-edge-states. However, the $D_{qm}$ is not localized at high symmetry lines but distributed across generic points in the BZ (Fig. 2(f)). The nonlinear Hall conductivity for the $\beta$ and $\gamma$ peaks with different in-plane spin orientations is provided in the Supplementary material.

\section{Doping effects}

Apart from EuSn$_2$As$_2$, we also examine its counterpart, EuSn$_2$P$_2$. Figure 4(c) illustrates the non-linear Hall conductivity $\sigma^{yxx}$ of EuSn$_2$As$_2$ and EuSn$_2$P$_2$ with a 4-layer slab geometry. Notably, we observe that the intensity of $\sigma^{yxx}$ in EuSn$_2$P$_2$ is considerably lower than that in EuSn$_2$As$_2$ near the Fermi level. In Fig. 4(a), the band structure of EuSn$_2$P$_2$ with a 4-layer slab geometry is depicted. Here, we discern a surface Dirac band with an energy gap of approximately 10 meV in EuSn$_2$P$_2$, whereas it remains nearly gapless in EuSn$_2$As$_2$ at the same thickness. Moreover, the band-edge-states of EuSn$_2$P$_2$ are significantly distant from the Fermi level. These band characteristics hinder EuSn$_2$P$_2$ from generating a substantial quantum metric dipole $D_{qm}$, thereby reducing the intrinsic non-linear Hall conductivity.

Given the significant impact of energy band characteristics on INHE, we propose P-doped EuSn$_2$As$_2$, denoted as EuSn$_2$As$_{2(1-x)}$P$_{2x}$, as an experimental platform for controlling INHE. Specifically, we calculate $\sigma^{yxx}$ of EuSn$_2$As$_{2(1-x)}$P$_{2x}$ using a linear interpolation of tight-binding model matrix elements of EuSn$_2$As$_2$ and EuSn$_2$P$_2$. Since tight-binding parameters encapsulate crucial information such as lattice constants and atomic bonding strength, interpolation is expected to capture all systematic changes in the electronic structure between the two endpoints. Thus, this approach proves highly reliable and effective for studying iso-valence substitution/doping ~\cite{RN195}. The calculated $\sigma^{yxx}$ with different doping ratios is presented in Fig. 4(d). Interestingly, we observe that the intensity of the $\alpha$ peak monotonically decreases with P doping, while the behavior of the $\beta$ peak is more intricate. The intensity of the $\beta$ peak gradually increases before reaching 20$\%$ doping. At 30$\%$ doping, $\sigma^{yxx}$ surpasses 40 mA/V$^2$, more than twice that of the undoped case. However, upon exceeding 40$\%$ doping, the intensity of $\sigma^{yxx}$ decreases rapidly. These findings underscore the effectiveness of doping as a means to control INHE in EuSn$_2$As$_2$.


\section{Discussion and Conclusion}

Nonreciprocal transport within semiconductor $p$-$n$ junctions constitutes a cornerstone of contemporary microelectronics, pivotal in the operation of diodes and transistors. However, traditional semiconductor junctions, particularly diodes, confront inherent limitations at high frequencies and low input voltages. Recent research underscores the potential of quantum geometric effects as a fundamentally novel operational principle for diodes. Consequently, materials exhibiting substantial intrinsic nonreciprocity can serve as diodes and are termed diodic quantum materials ~\cite{RN178}, ~\cite{TRC_1}, ~\cite{TRC_2}, ~\cite{TRC_3}. By virtue of their junction-free nature, diodic quantum materials transcend the constraints associated with semiconductor junctions, functioning as high-efficiency rectifiers capable of operating at terahertz frequencies, zero bias voltage, and ultralow input power. Thus, the quest for quantum materials offering a new platform for quantum phenomena and technology is of paramount importance. 

Our study demonstrates that the AFM topological material EuSn$_2$As$_2$, with even-numbered layer slab geometry, exhibits significant non-linear Hall conductivity $\sigma^{yxx}$, stemming from Type-II surface Dirac band dispersion and anti-crossing band gaps of band-edge-states. Moreover, appropriate P doping can substantially enhance the value of $\sigma^{yxx}$. Given the direct relationship between device application efficiency and non-linear Hall voltage, the large non-linear Hall conductivity observed in the EuSn$_2$As$_2$ thin-film holds promise for improving device operating efficiency. An exciting future breakthrough would be to demonstrate room-temperature wireless rectification based on the INHE in an AFM material.

Beyond the A-type AFM configuration explored in this work, recent studies propose that NLHE can also manifest in more intricate magnetic configurations, such as altermagnetic materials ~\cite{TRC_6} and non-collinear antiferromagnetism ~\cite{TRC_7}. In addition to intrinsic effects arising from the band structure, disorder plays a crucial role in inducing NLHE in PT-symmetric antiferromagnetic materials ~\cite{TRC_8}, ~\cite{TRC_9}, ~\cite{TRC_10}, ~\cite{TRC_11}, ~\cite{TRC_12}. Intriguingly, this extrinsic effect can surpass intrinsic contributions in magnitude, emphasizing the need for further theoretical investigations to clarify its impact on real materials. Exploring the interplay between these intrinsic and extrinsic mechanisms, as well as their implications for potential spintronics applications, remains an open and exciting avenue for future research.

\section{Acknowledgement}
T.-R.C. was supported by National Science and Technology Council (NSTC) in Taiwan (Program No. MOST111-2628-M-006-003-MY3 and NSTC113-2124-M-006-009-MY3), National Cheng Kung University (NCKU), Taiwan, and National Center for Theoretical Sciences, Taiwan. This research was supported, in part, by the Higher Education Sprout Project, Ministry of Education to the Headquarters of University Advancement at NCKU. T.-R.C. thanks the National Center for High-performance Computing (NCHC) of National Applied Research Laboratories (NARLabs) in Taiwan for providing computational and storage resources. H.L. acknowledges the support by Academia Sinica in Taiwan under grant number AS-iMATE-113-15. L.F. was supported by the Air Force Office of Scientific Research under award num ber FA2386-24-1-4043. The Global Seed Funds provided by MIT International Science and Technology Initiatives (MISTI) under the program of MIT Greater China Fund for Innovation is acknowledged.

\begin{figure}[b]
    \centering
    \includegraphics[width=15cm]{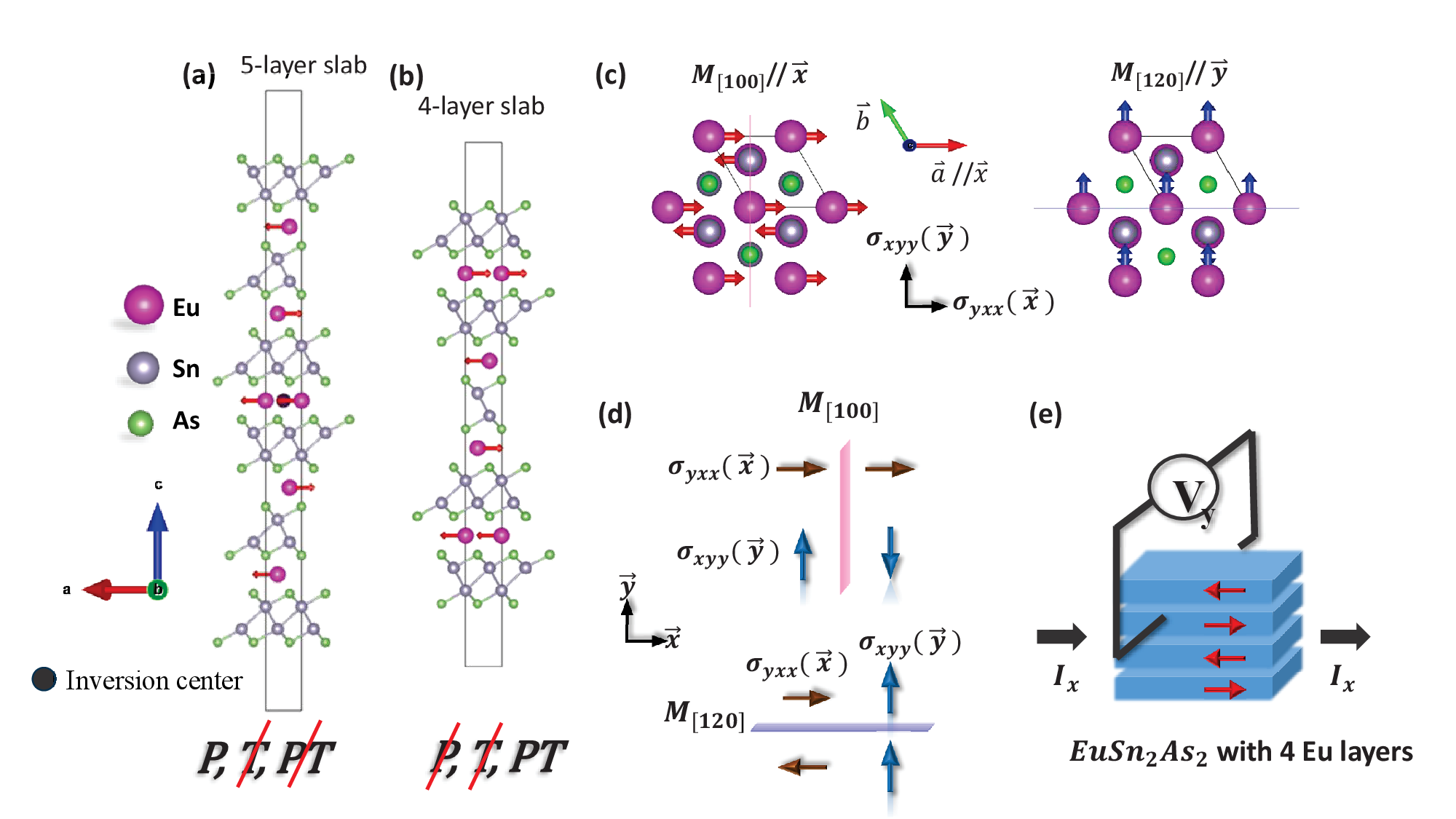}
    \caption
    {
    (a) The crystal structure of bulk EuSn$_2$As$_2$ with  5-layer slab model, where the black solid dot represents the inversion center. The red arrows represent the spin moments along the $x$-axis.
    (b) The 4-layer slab model.
    (c) Left panel: The mirror plane $M_{[100]}$ of EuSn$_2$As$_2$ when the spin magnetic moment is along the $x$-axis. Right panel: The mirror plane $M_{[120]}$ of EuSn$_2$As$_2$ when the spin magnetic moment is along the $y$-axis.
    (d) Illustration depicting the transformation of the pseudovector of in-plane non-linear conductivity under the $M_{[100]}$ and $M_{[120]}$ mirror plane.
    (e) Illustration showcasing the INHE in EuSn$_2$As$_2$ with a 4-layer slab model.
    } 
    \label{Fig1}
\end{figure}

\begin{figure}[b]
    \centering
    \includegraphics[width=15cm]{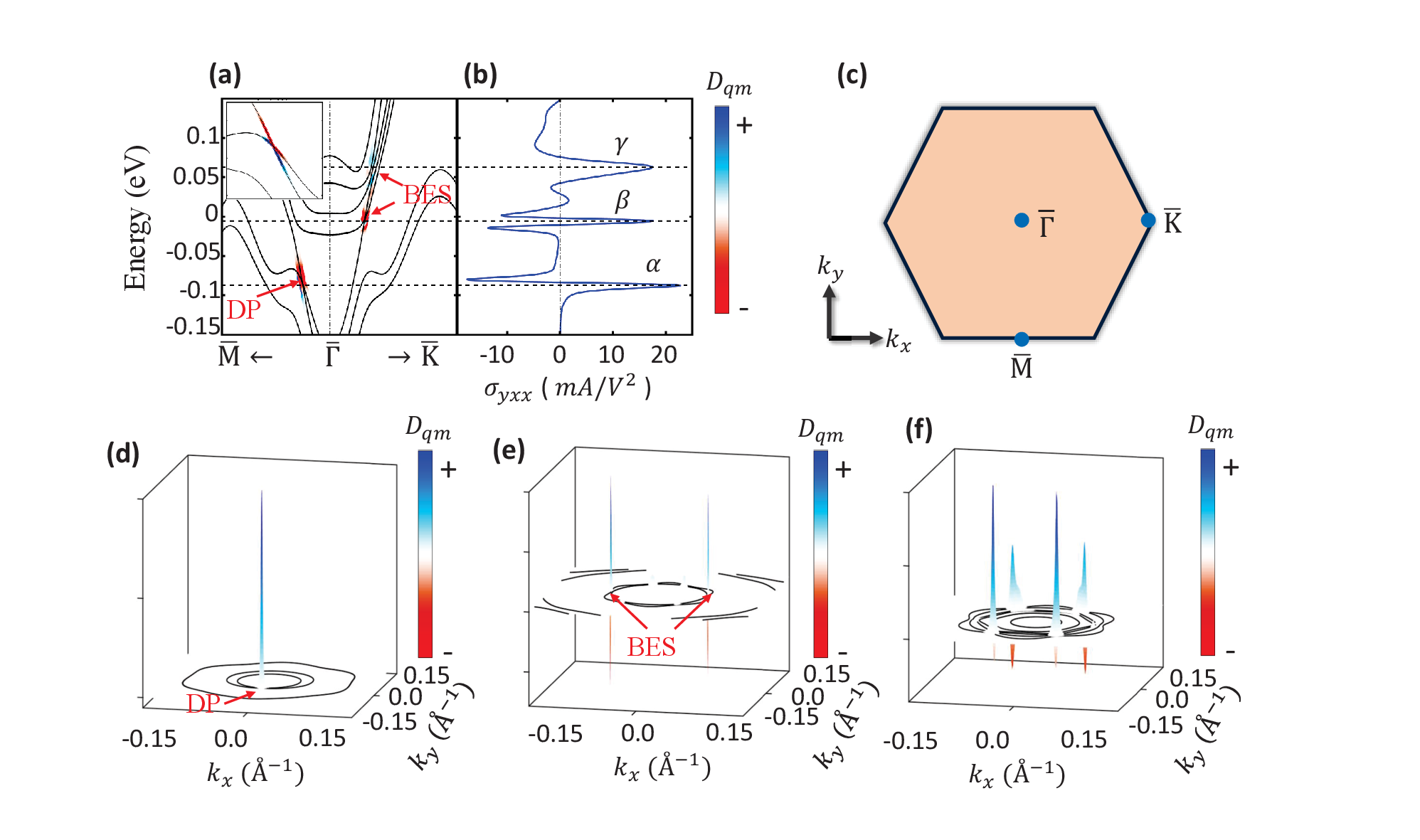}
    \caption
    {
    (a) The band structure of a 4-layer slab of EuSn$_2$As$_2$ with A-type antiferromagnetism. The color gradient represents the strength of quantum metric dipole $D_{qm}=v_yg_{xx}-v_xg_{yx}$.
    (b) The calculated non-linear Hall conductivity $\sigma^{yxx}$ under varying chemical potentials at 20K.
    (c) The Brillouin zone (BZ).
    (d-f) The distribution of $D_{qm}$ across the BZ, corresponding to peaks $\alpha$, $\beta$, and $\gamma$, respectively.
    }
    \label{Fig2}
\end{figure}

 \begin{figure*}[htbp]
     \centering
     \includegraphics[width=12cm]{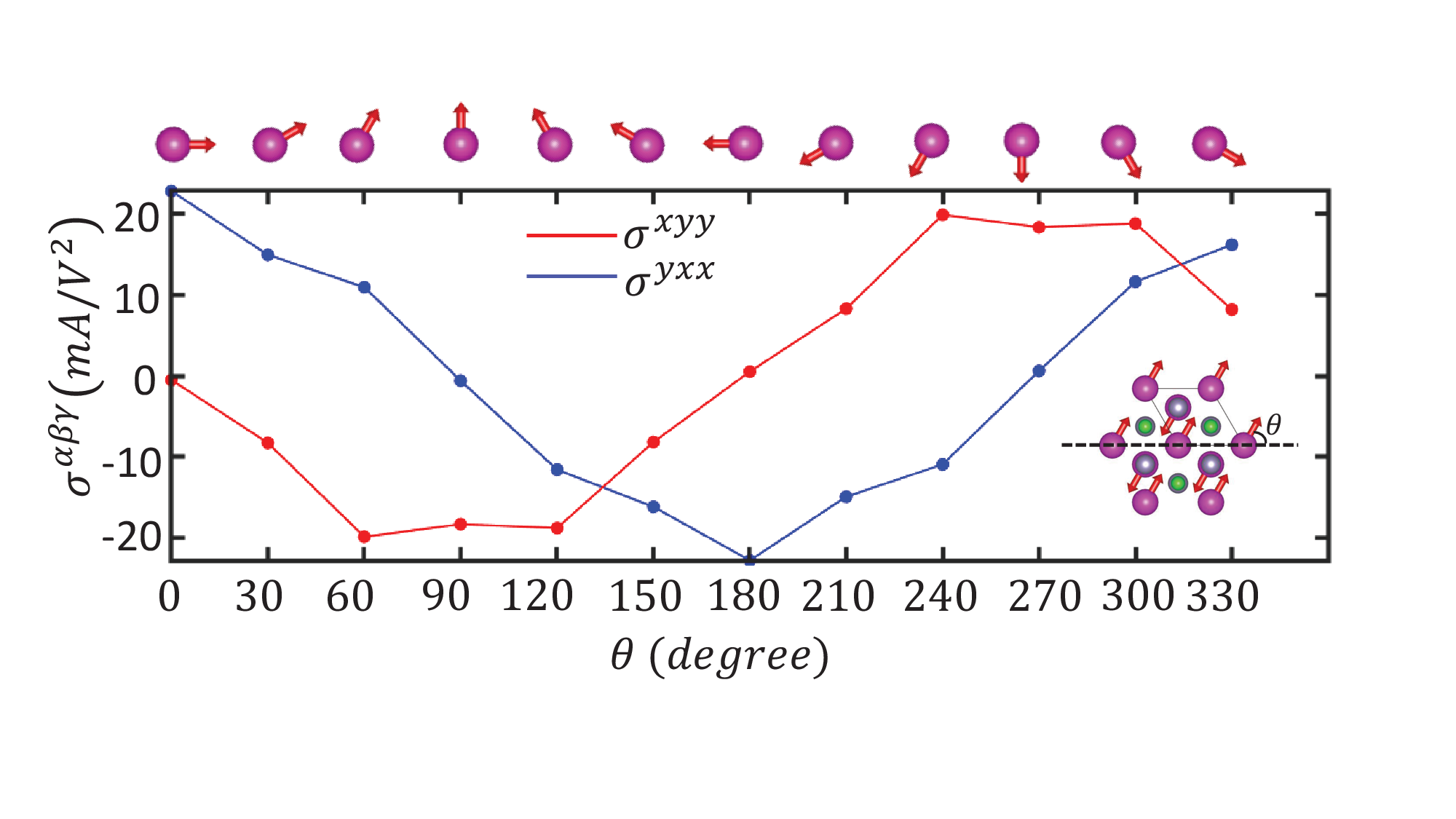}
    
     \caption{
         The calculated non-linear Hall conductivity of a 4-layer slab of EuSn$_2$As$_2$ with spin moment rotation in the $xy$ plane. Here, the red line and blue dot represent $\sigma^{xyy}$ and $\sigma^{yxx}$, respectively.
             }
     \label{Fig3}

 \end{figure*}

\begin{figure*}[b]
   \centering
   \includegraphics[width=15cm]{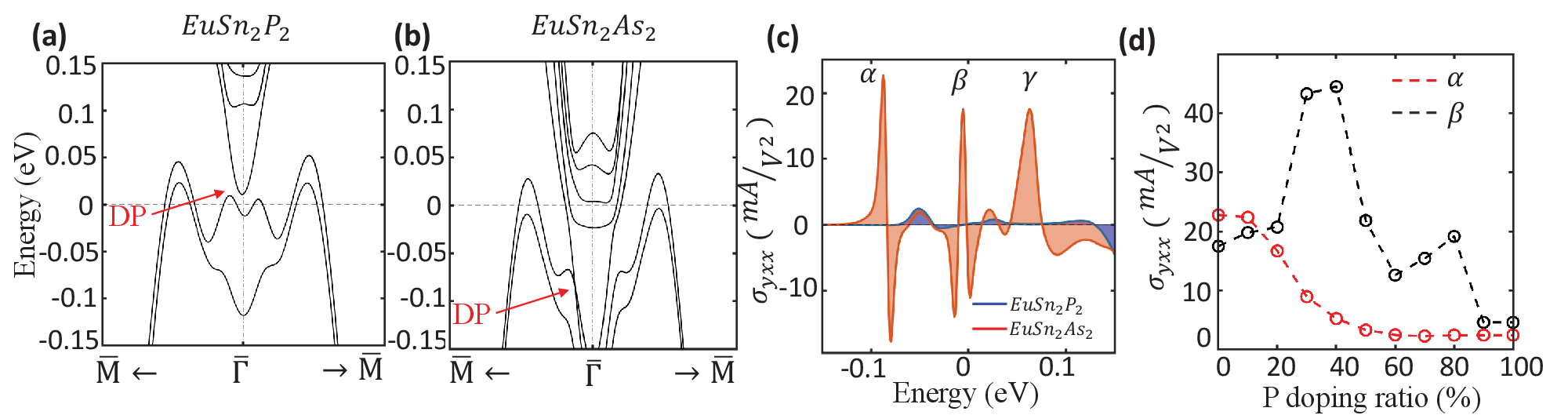}
   \caption{
    (a-b) Band structures of 4-layer slab models for EuSn$_2$P$_2$ and EuSn$_2$As$_2$, respectively.
    (c) Comparison of $\sigma^{yxx}$ between EuSn$_2$P$_2$ (blue line) and EuSn$_2$As$_2$ (red line) at 20K.
    (d) Variation of $\sigma^{yxx}$ for peaks $\alpha$ (red circle) and $\beta$ (black circle) at 20K under different P doping ratios.
           }
   \label{Fig5}
\end{figure*}

\bibliographystyle{apsrev4-2}
\hypersetup{
                colorlinks = true,
                linkcolor  = blue,
                filecolor  = magenta,
                urlcolor   = blue,
                citecolor  = blue
            }



\end{document}